\documentclass[gmd, manuscript]{copernicus}

\begin{document}

\title{Parcels v0.9: prototyping a Lagrangian Ocean Analysis framework for the petascale age}

\Author[1]{Michael}{Lange}
\Author[2,3]{Erik}{van Sebille}

\affil[1]{Grantham Institute \& Department of Earth Science and Engineering, Imperial College London, UK}
\affil[2]{Institute for Marine and Atmospheric research Utrecht, Utrecht University, Utrecht, Netherlands}
\affil[3]{Grantham Institute \& Department of Physics, Imperial College London, UK}

\runningtitle{PARCELS: LAGRANGIAN OCEAN ANALYSIS}
\runningauthor{LANGE AND VAN SEBILLE}
\correspondence{Erik van Sebille (e.vansebille@uu.nl)}

\received{}
\pubdiscuss{} %% only important for two-stage journals
\revised{}
\accepted{}
\published{}

%% These dates will be inserted by Copernicus Publications during the typesetting process.

\firstpage{1}

\maketitle

\nolinenumbers

\begin{abstract}

As Ocean General Circulation Models (OGCMs) move into the petascale age, where the output of single simulations exceeds petabytes of storage space, tools to analyse the output of these models will need to scale up
too. Lagrangian Ocean Analysis, where virtual particles are tracked through
hydrodynamic fields, is an increasingly popular way to analyse OGCM output, by
mapping pathways and connectivity of biotic and abiotic particulates. However,
the current software stack of Lagrangian Ocean Analysis codes is not dynamic
enough to cope with the increasing complexity, scale and need for customisation
of use-cases. Furthermore, most community codes are developed for stand-alone
use, making it a nontrivial task to integrate virtual particles at runtime of
the OGCM. Here, we introduce the new Parcels code, which was designed from the
ground up to be sufficiently scalable to cope with petascale computing. We
highlight its API design that combines flexibility and customisation with the
ability to optimise for HPC workflows, following the paradigm of domain-specific
languages. Parcels is primarily written in Python, utilising the wide range of
tools available in the scientific Python ecosystem, while generating low-level
C-code and using Just-In-Time compilation for performance-critical computation.
We show a worked-out example of its API, and validate the accuracy of the code
against seven idealised test cases. This version~0.9 of Parcels is focussed on
laying out the API, with future work concentrating on support for curvilinear grids,
optimisation, efficiency and at-runtime coupling with OGCMs.

\end{abstract}

\introduction

Lagrangian Ocean Analysis, where virtual particles are tracked within the flow
field of hydrodynamic models, has over the last two decades increasingly been
used by physical oceanographers and marine biologists alike
\citep{vansebille2017}.The particles can represent passive parcels of seawater \citep[e.g.][]{doos1995, blanke1997}
or its constituent tracers such as nutrients \citep[e.g.][]{Jonsson2011,
Qin2016}, as well as particulate matter such as microbes
\citep[e.g.][]{Hellweger:2014gb, Doblin:2016ig}, larvae
\citep[e.g.][]{cowen2006,paris2005,Teske:2015jk,CetinaHeredia:2015gq}, pumice
\citep[e.g.][]{Jutzeler:2014fq}, plastic litter \citep[e.g.][]{lebreton2012}, or
icebergs \citep[e.g.][]{Marsh:2015dn}. The trajectories of the virtual particles
can be used to analyse the flow within Ocean General Circulation Models (OGCMs)
and other velocity fields for dispersion characteristics
\citep[e.g.][]{BeronVera:2015wn}, Lagrangian Coherent Structures
\citep[e.g.][]{Haller2015}, water mass pathways and transit times
\citep[e.g.][]{Ruhs2013}, Lagrangian streamfunctions \citep[e.g.][]{Doos2008a}
and biological connectivity between regions \citep[e.g.][]{kool2013}. See
\citet{vansebille2017} for an extensive review on Lagrangian Ocean Analysis.

There are currently three main community codes available to calculate the
trajectories of virtual particles in Ocean General Circulation Models: Ariane
\citep{blanke1997}, TRACMASS \citep{doos:2013, doos:2017}, and the Connectivity
Modeling System \citep[CMS,][]{Paris:2013fs}. These codes, being open-source and
having excellent support teams, have served the wider community very well over
the past decades. However, it is not clear that these three codes will be able
to scale up easily to the petascale age of computing, where particle trajectory
codes will need to be able to deal with potentially petabytes of hydrodynamic
field data and gigabytes of particle trajectory data. Exploring
advanced optimization strategies to overcome these big-data challenges, such
as coupled (online) execution with the host OGCM or reducing the
volume of hydrodynamic data by selectively filtering data regions based
on particle locations, will require a flexible execution model that
can dynamically be adapted to complement the respective data and
execution formats of various host OGCMs.

Furthermore, the current stack of codes is mostly built for the tracking of
water parcels or passive particulates. While the CMS and TRACMASS do support the
addition of diffusive processes through Markovian stochastic models
\citep[e.g.][]{griffa1996applications}, it is non-trivial to incorporate
`behaviour' of particulates to these codes. Effortless incorporation of
behaviour such as sinking, fragmentation, or even swimming to particulates would
simplify exploration of the dynamics of particulates such as fish, icebergs and
marine debris.

Here, we describe a novel framework for computing Lagrangian particle
trajectories, named Parcels (`Probably A Really Computationally Efficient
Lagrangian Simulator'). Being developed from the ground up with scalability and
performance in mind, we hope that this Parcels framework will be able to keep up
with OGCM development for the coming decades, particularly by being scalable and efficient at reading in hydrodynamic data. We have furthermore focussed on flexibility
and customisability of the particle dispersion schemes, so that it is relatively
straightforward to add new functionality such as active particle behaviours.

We have decided to brand this version of Parcels as v0.9, signalling that while
in principle it is feature-complete, the code is not nearly as fast and efficient
as we envision it to be in the future. Improving performance will be the main
priority as we work towards v1.0. We invite all interested researchers to
contribute to the development by starting to use the code.

While development efforts of Parcels focus on oceanographic
applications, the Parcels framework  should in principle also be
adaptable to atmospheric particle tracking simulations. Models such as
FLEXPART \citep{Stohl:2005wca} and the MetOffice NAME model
\citep{Jones2007} are state-of-the-art and have an excellent
track-record in the field of atmospheric dispersion modelling, but
perhaps some of the ideas presented here could be incorporated or used
in these models too.

This paper is structured as follows: in the next section, we will describe the
philosophy behind the Parcels code. We then present a worked-out example of an
application of Parcels for an actual scientific experiment in
Sect.~\ref{sec:example}. Following that, we evaluate the accuracy of the code in
Sect.~\ref{sec:eval}, by comparison to analytical solutions in idealised test.
We provide a future outlook in Sect.~\ref{sec:outlook}, before concluding in
Sect.~\ref{sec:conclusions}.

\section{Prototype design and philosophy}

A key contribution of the new Parcels v0.9 framework is to
define a set of interfaces and composable abstractions that encapsulate
the various processes required to create dynamic and extensible
Lagrangian models that feature direct interactions between particles
and an associated OGCM grid. The design follows modern scientific software
engineering practices, providing high levels of modularity and
flexibility with a clear intent to further specialize various
sub-components at a later stage. The interfaces provided in Parcels
are therefore intended to capture the general domain-specific
challenges posed by particle tracking for Lagrangian Ocean
Analysis. The overall design philosophy, as well as the structure of
the code, are driven by three major design considerations:

\begin{itemize}
\item {\bf Extensibility} -- While the core algorithm of Lagrangian
  particle models is concerned with the advection and dispersion of
  passive particles that constitute infinitely small point parcels,
  practical oceanographic applications often require more complex
  behaviour of the particles. Potential extensions towards
  individual-based modelling of particulates to simulate biological
  species or marine debris will require extensions to particle data
  definitions and programmable behavioural customisation at a
  per-particle level.

\item {\bf Compatibility} -- Particle tracking in oceanography
  requires the close coupling of computational particles to velocity
  data that defines the hydrodynamic flow field. Parcels aims to make as little
  assumptions about the nature and structure of the hydrodynamic fields as
  possible, so as to be compatible with various types of OGCMs and
  data formats. While the focus in this v0.9 is on utilising offline
  data, this includes considerations for at-runtime coupling with
  OGCMs in the future.

\item {\bf Dynamic data} -- Particle data is sparse in nature and can,
  depending on application context, exhibit very dynamic data access
  patterns where new particles are inserted and deleted from the
  active set at runtime. For this reason, structured compile-time
  performance optimisations and parallelisation strategies are
  insufficient, and Just-In-Time scheduling is required to handle the
  amorphous data parallelism inherent in dynamic particle
  applications~\citep{Pingali2011}.
\end{itemize}

The above list of requirements suggests that a static compile-time
approach is likely to provide insufficient flexibility to adjust to the
various scientific contexts in which oceanographic particle tracking
might be utilised. For this reason Parcels is based on the
domain-specific languages paradigm, which aims to decouple the problem
definition as defined by the scientific modeller from the implementation
that is ultimately executed on a particular hardware
architecture. This approach is based on automated code generation at
runtime and creates a separation of concerns between domain scientists
and computational experts that allows hardware-specific performance
optimisation and thus greater flexibility with respect to advances in
high-performance computing resources.

Since the prototype of the Parcels framework presented here provides a
conceptual blueprint for future versions, we define a clear set of
abstractions for the following three software layers:

\begin{itemize} \item {\bf User-facing API} -- The primary objective of
Parcels is to provide a user-friendly, clear and concise API for
scientists to perform oceanographic particle tracking experiments with
very little effort, while leaving room for customisations that go beyond
traditional configuration files. For this reason Parcels provides a
high-level Python API that enables users to define a complete model in
a small number of lines of code (see examples in
Sect.~\ref{sec:example}). For more advanced models, the API
also provides enough scope to fully control the variable layout of
particles in memory, as well as to define custom behaviour via
individual kernel operations.

\item {\bf Execution layer} -- The transient nature of Lagrangian
particles implies that many practical oceanographic applications rely on
particle sets that may grow and shrink dynamically, while also relying
on external hydrodynamic field data that might be sampled at a timestep
much different from the primary particle loop. This complex parameter
variability entails that the core loop that updates individual particle states
needs to be highly dynamic and flexible, as well as highly optimised for
large-scale applications. Parcels aims to encapsulate the core
parameters of the particle update loop so as to establish an interface
for integration with a variety of external host OGCMs, and leaves enough
scope for more advanced performance optimisations in the future.

\item {\bf Data layout} -- The two fundamental types of data involved in
Lagrangian particle tracking algorithms constitute field data provided
by the external OGCM, as well as data on the particle state. Since the
data layout for particle data might change with future performance
optimisations, and the memory layout of field data depends on the OGCM
implementation, Parcels provides high-level abstractions for both types
of data, allowing the actual data layout in memory to change.

\end{itemize}

\begin{figure}[t] \centering
  \includegraphics[width=\linewidth]{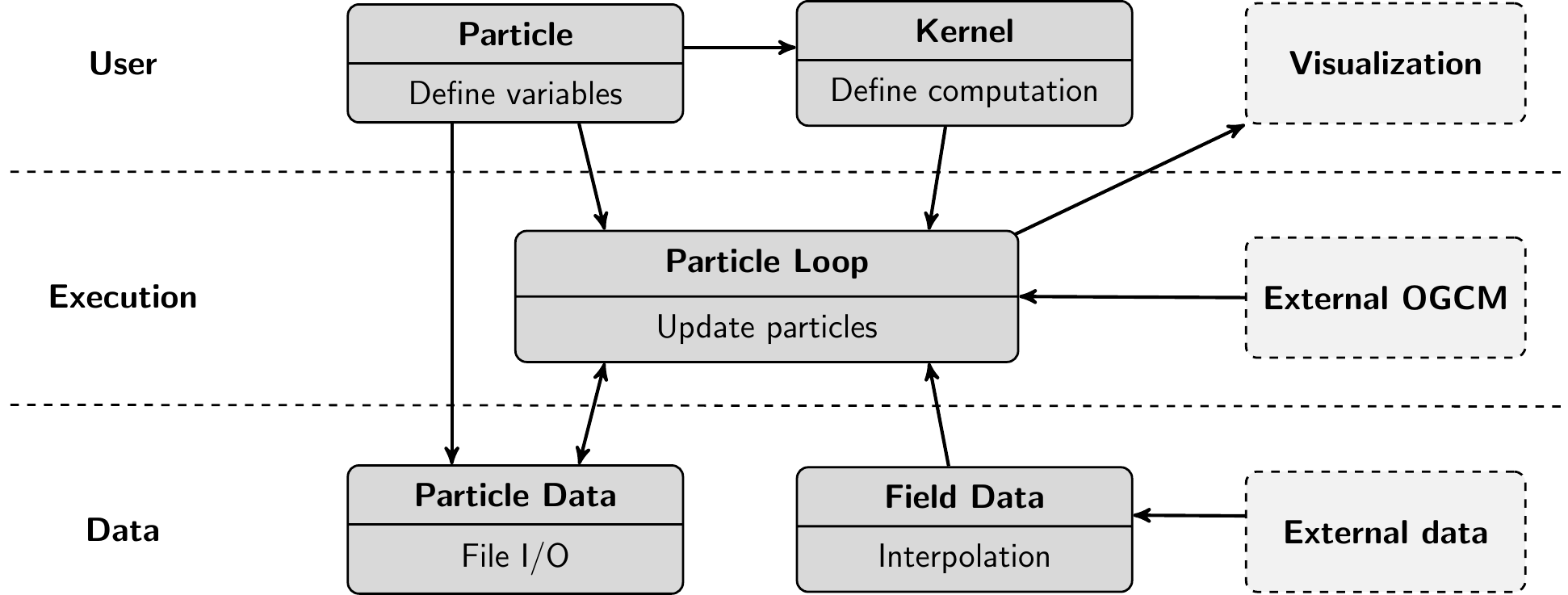}
  \caption{Conceptual abstractions (dark) and functionalities
    encapsulated in the Parcels prototype in relation to external
    components (light).}
  \label{fig:parcels_layers}
\end{figure}

The abstractions shown in Fig.~\ref{fig:parcels_layers} comprise the
core functionalities provided by the framework. The primary input in the
user layer consists of generic definitions of the particle variables for
individual types of particles, alongside an interface to define the
computation kernels. Parcels' core execution loop uses this information
to update particle data given external parameters, such as timestepping
constraints, and interpolated hydrodynamic field data. Thus, given a stable
user-level API and a highly modular code structure, it is possible to
implement various applications and experiments without commiting to a
particular implementation, while leaving enough scope for further
development and future performance optimisation `under the hood'.

\subsection{Programmable user interface}

The prototype presented in this paper provides a highly flexible user
API that allows users to define complete models via the Python
programming language. The user hereby manages creation, execution and
customisation of individual sets of particles, as well as combinations
of computational kernels to update the particle state. In contrast to
traditional configuration files, this approach provides the user with
native compatability with the open-source libraries and tools available
in the scientific Python ecosystem.

\begin{figure}[t] \centering
  \includegraphics[width=\linewidth]{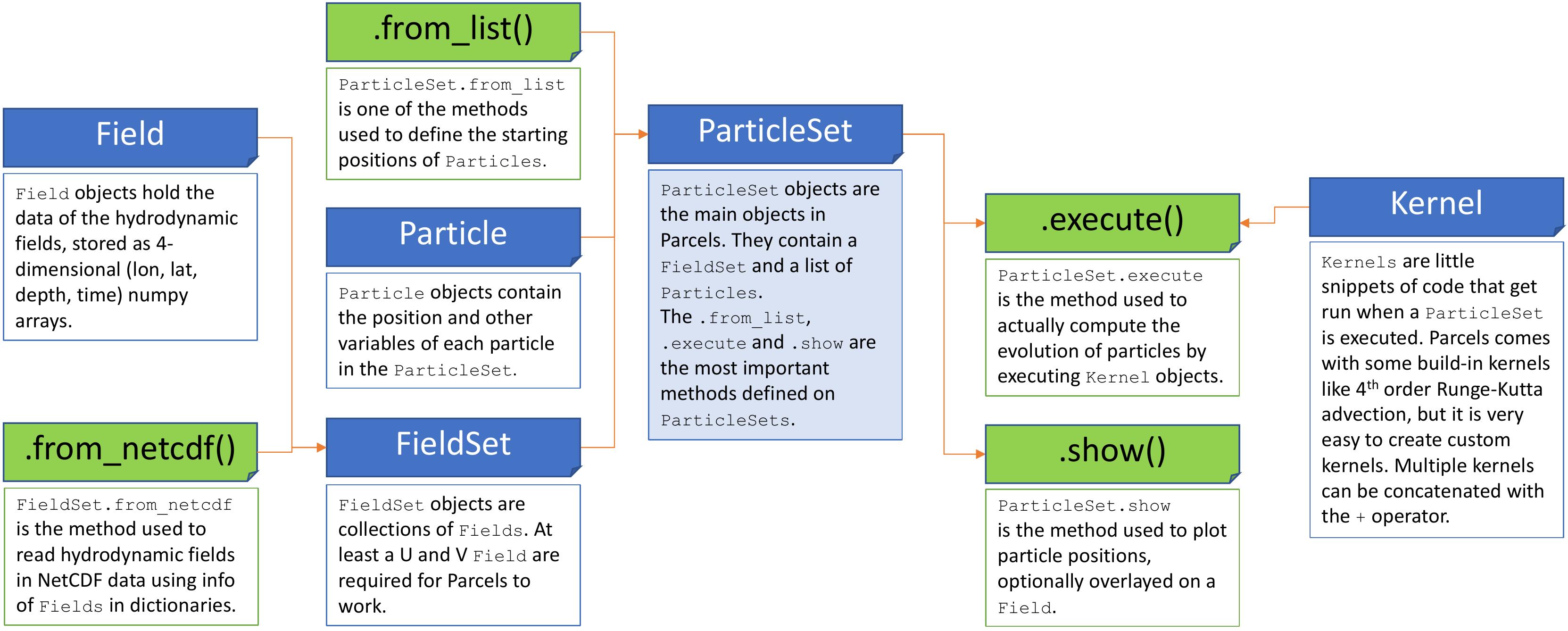}
  \caption{Class diagram of the Parcels v0.9 prototype implementation. Classes are depicted in blue, methods in green. Note that not all methods and classes are shown in this diagram.}
  \label{fig:parcels_design}
\end{figure}

The key components of Parcels' overall class structure are depicted in
Fig.~\ref{fig:parcels_design}. The definition of the variables that
constitute a single particle is hereby encapsulated in the
\texttt{Particle} class, while container objects of type
\texttt{ParticleSet} provide the runtime handling and management of
particle data. Python \emph{descriptor} objects are used to generically
define the compound data type underlying each type of particle, leaving
allocation and memory layout choices to the particular implementation of
the data container structure.

The computational behaviour of particles is encapsulated through the
\texttt{Kernel}. Parcels provides a set of pre-defined advection
methods, as well as allowing users to define custom behaviour
programmatically. Multiple kernels can be concatenated, allowing users
to incrementally build complex behaviour from individual components.

\subsubsection{Advection algorithm}

At its core, computing Lagrangian particle trajectories is equivalent to
solving the following equation: 
\begin{equation} 
\boldsymbol{X}(t+\Delta t) = \boldsymbol{X}(t) +
\int\limits_{t}^{t+\Delta t} \boldsymbol{v}(\boldsymbol{x},\tau) \,
\mathrm{d}\tau + \Delta \boldsymbol{X}_b(t), 
\label{eq:trajectory-equation} 
\end{equation} 
where $\boldsymbol{X}$ is the three-dimensional position of a particle,
$\boldsymbol{v}(\boldsymbol{x},t)$ is the three-dimensional velocity
field at that location from an OGCM, and $\Delta\boldsymbol{X}_b(t)$ is
a change in position due to `behaviour'. The latter can itself be an
integration of a (three-dimensional) velocity field, for example when a
particle sinks downward because of a negative buoyancy force.

In Parcels, the trajectory equation (\ref{eq:trajectory-equation}) is by
default time-stepped using a 4$^{th}$ order Runge-Kutta scheme, although
schemes for Euler-Forward and adaptive Runge-Kutta-Fehlberg integration
\citep[RKF45, e.g.][]{alexander1990solving} are also provided. In
principle, the Parcels framework should be flexible enough to also implement
integration using the discrete analytical streamtube method
\citep{blanke1997, doos:2017}.

\subsubsection{Custom kernels}

Lagrangian particle tracking in the ocean often involves more complex
displacement schemes than simple velocity-driven advection. For example,
in the presence of turbulence, a Random Walk kernel or Brownian motion
is required, while ocean ecology models often include active locomotion. Parcels therefore
allows users to create generic kernel functions by providing native
Python functions that adhere to the function signature
\texttt{KernelName(particle, fieldset, time, dt)}. Within these kernel
functions, users can access built-in particle state variables,
such as \texttt{particle.lat} and \texttt{particle.lon}, or user-defined
ones. Access to field data from within kernels is provided through the
\texttt{fieldset} object, which provides fields as named properties,
for example \texttt{fieldset.U} for the zonal velocity. Interpolation
of field data is implemented via overloaded member access on the field
object (square bracket notation), allowing user to express field
sampling as \texttt{fieldset.fieldname[time, lon, lat, depth]}.

In addition to kernels that update the internal state of particles,
Parcels' execution engine also enables users to customize the
behaviour of particles under various error conditions. For this, a
similar type of kernel function can be created and passed to the
execution call, mapped to a particular error type that might be triggered
during the main particle update, for example
\texttt{OutOfBoundsError}.

\subsection{Execution and JIT compilation}

The update of the internal state of particles is facilitated by a
dynamic loop, which applies a user-defined combination of kernels to
each particle in a \texttt{ParticleSet}. The primary particle update
loop can either be run with a forward timestepping, or in a
time-backward mode to enable inverse modelling. For this central
update loop, Parcels provides two modes of execution:

\begin{itemize}
\item \textbf{Scipy mode:} A pure Python mode that utilises
  \texttt{interpolator} objects provided by the Scientific Python
  package (SciPy) to perform interpolation of field data. This mode is
  primarily intended as a debug option due to the performance penalty
  of running kernels in the Python interpreter itself.

\item \textbf{JIT mode:} Runtime code generation and Just-In-Time
  compilation (JIT) are utilised to generate low-level C code that
  performs the particle state update and field data interpolation. The
  code generation engine hereby primarily translates a restricted
  subset of the Python language into equivalent C code, while a set of
  utility modules provides auxiliary functionality such as random
  number generation or mathematical utilities (\texttt{math.h}).
\end{itemize}

The execution mode of the particle update loop is determined by the type
of the particle (\texttt{ScipyParticle} or \texttt{JITParticle}) used to
create the \texttt{ParticleSet}. Development of new features in the
current Parcels prototype is strongly driven by the fact that both modes
are intended to be semantically equivalent. This means that new
features can rapidly be developed using the full flexibility of the
Python interpreter, providing a template implementation and test case
for implementation in the computationally more efficient JIT
mode.

Parcels' dynamic update loop also provides an \texttt{interval} keyword
to impose a secondary sub-timestepping that allows for direct coupling
with a host OGCM in the future. The dynamic composition of multiple
timestepping intervals might also be used for future data and
performance optimisation strategies, for example directed prefetching
of regional field data. Such strategies, as well as a potentially more
intricate execution engine, have to be explored carefully to
successfully tackle the big-data challenges facing Lagrangian tracking
codes in the petascale age.

\subsection{Interpolation}

The interaction of particles with their enclosing fields is currently
limited to interpolating field data onto the current particle position.
In the SciPy debug mode this is facilitated by
\verb|scipy.interpolate.RegularGridInterpolator| objects and supports linear
and nearest-neighbour interpolation. Equivalent low-level C routines are
also included in the Parcels source code as macros that can be inlined
into the generated C kernel code by the code generation engine. More
advanced interpolation methods, such as quadratic, cubic or spline
interpolation, may easily be added in future releases if a fast
C implementation can be provided with Parcels' internal header files.

One of key performance advantages of using runtime code generation is
the ability to inline bespoke grid interpolation methods with the
user-defined kernels in Parcels to avoid the Python interpreter
overhead of repeatedly calling native Python interpolation functions.
This overhead can be quite significant due to the high frequency at
which the associated field data needs to be sampled.  This can be
illustrated using the ``Steady-state flow around a peninsula'' test case discussed
in Sect.~\ref{sec:eval-peninsula}, where $100$ particles are advected
for $20$ hours with a timestep size of $30$ seconds. While the
sequential execution time of the pure Python implementation runs in
$305.92$ seconds, the auto-generated JIT kernels can run the same
experiment in $1.74$ seconds, a speedup of over $150\times$.

\subsection{External field data}

Parcels v0.9 supports external field data from
NetCDF files, with a configurable interface to describe the input data
and variable structure. The data is encapsulated in individual
\texttt{Field} objects, which are accessible from within particle
kernels via provided interpolation routines. Individual fields are
stored in a \texttt{FieldSet} container class, which may also provide
global meta-data to the kernel execution engine at runtime.

Currently, only linear interpolation schemes are implemented in Parcels,
both in space and in time. In space, Parcels can currently only work on
regular grids (i.e. where the grid dimensions are functions of only
longitude, only latitude or only depth). However, support for
unstructured grids is a priority for the next release of the code,
Parcels v1.0.

\section{A worked-out example: tracking virtual foraminifera in the Agulhas region}\label{sec:example}

To highlight some of the prototype design and philosophies of the Parcels API, we
here present a worked-out example code of a previously-published scientific
experiment. This example follows the experimental design of
\citet{vanSebille:2015ed}, where the goal was to investigate the temperatures that
planktic foraminifera experience during their lifespan as they drift with the
currents in the upper ocean. In particular, that study looked at the variability
of lifespan-averaged temperatures of foraminifera that all end up on one single
location on the ocean floor \citep[e.g.][]{peeters2004, Katz:2010vv}.

Figure 1b of \citet{vanSebille:2015ed} depicted the origin of virtual planktic
foraminifera that end up on a site just off the coast of Cape Town
(17.3$^\circ$E, 34.7$^\circ$S), at 2,440 meter water depth. The virtual particles were
released at that site and then tracked in time-backward mode. There were two
phases to the experiment: in the sinking phase, the foraminifera were tracked
back as they sunk at 200 meter per day to the ocean floor, while being advected
by the (deep) ocean circulation. In the lifespan phase, the particles were
then tracked further backward in time as they were advected by the horizontal
circulation at their 50m dwelling depth. During this last phase, temperature
along their trajectory was recorded at daily interval.

While the original experiment was computed with the Connectivity Modelling
System \citep{Paris:2013fs}, here we have re-coded it using the Parcels API.
This experiment setup is a fitting one, as it combines a number of the API
highlights of Parcels: custom kernels, NetCDF I/O, and field sampling. The full
Python code for this experiment in Parcels is available at
\url{https://doi.org/10.5281/zenodo.823994}. Below, we emphasise some of the key
statements in the Python script.

\subsection{Reading the FieldSet}\label{sec:example_fieldset}

The hydrodynamic fields that carry the foraminifera come from the OFES
model \citep{masumoto2004} and can be accessed from
\url{http://apdrc.soest.hawaii.edu/datadoc/ofes/ncep\_0.1\_global\_3day.
php}. Three-dimensional velocities and temperature are available on
1/10$^\circ$ horizontal resolution, on 54 vertical levels, and are
stored as three-day averages. The bash script
\verb|get_ofesdata_agulhas.sh| provided at
\url{https://doi.org/10.5281/zenodo.823994} was used to download
snapshot numbers 3165 to 3289, covering the year 2006, in a subdomain
around the core site off Cape Town (note, the total file size is 6GB).

While the 6GB file size for this example is not excessively large and
could in principle be loaded into memory all at once, this will not be
possible for \verb|FieldSets| with larger regional domains or longer
time series. Hence, Parcels provides a system to read in hydrodynamic
fields during particle integration, at any time storing only three
consecutive timeslices \citep[e.g.][]{Paris:2013fs}. See also Section
\ref{sec:example_executing}.

After the first three days of hydrodynamic fields are read in through a
call to the user-defined \verb|set_ofes_fieldset| function (see the
\verb|example_corefootprintparticles.py| script for the exact
formulation of this function, which requires as input a 
set with filenames, provided as a list of arbitrary length), three
global constants are added to the \verb|FieldSet| 

\includegraphics[scale=1.0]{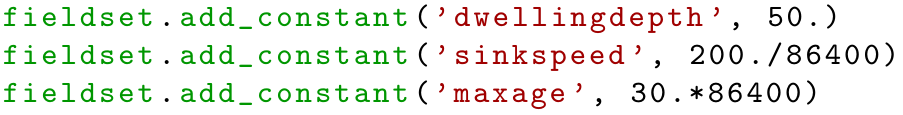}

These constants will be used later in the custom kernels controlling the
movement of the particles.

\subsection{Defining the ParticleSet}

Apart from information on their location and time, the virtual foraminifera
particles will need two extra \verb|Variables|: the sea water temperature at
their present location, and their age. Therefore, we define a new particle
class, which inherits from the standard \verb|JITParticle|: 

\includegraphics[scale=1.0]{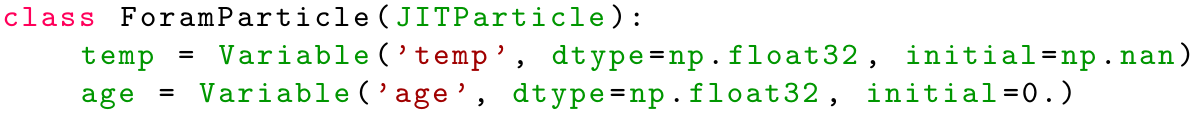}

And we then define a \texttt{ParticleSet} containing a single particle
as

\includegraphics[scale=1.0]{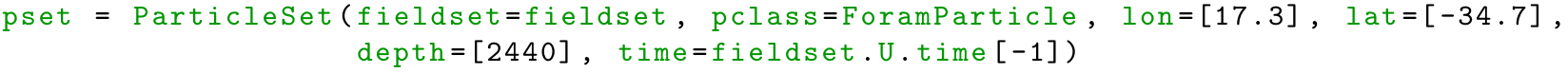}

\subsection{Defining the custom kernels}\label{sec:example_kernels}

We need to define four custom kernels: one that causes the particle to sink
after it dies, one that keeps track of its age and deletes it once it reaches
its maximum age, one that samples the temperature at its location, and
one that deletes the particle when it reaches a boundary of the domain (since we
only have hydrodynamic data in a subset of the global OFES domain). Note that
while in principle the first three could be written in one Kernel, here we write
three separate kernels and then concatenate these with the built-in
\verb|AdvectionRK4_3D| kernel.

The first kernel, controlling the sinking of the particle after it died (i.e.
the first twelve days in our reverse-time experiment), can be written as

\includegraphics[scale=1.0]{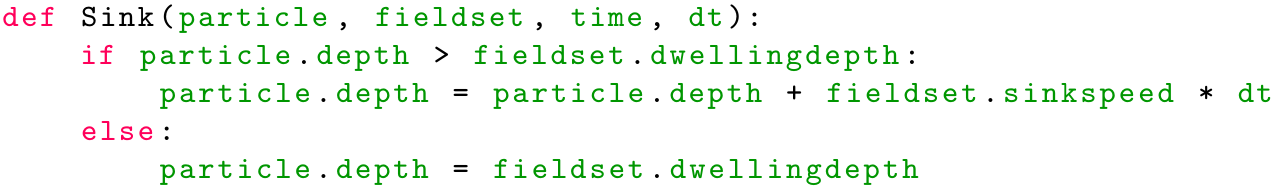}

The second kernel, which keeps track of the age and deletes the particle when it
reaches \verb|maxage|, can be written as

\includegraphics[scale=1.0]{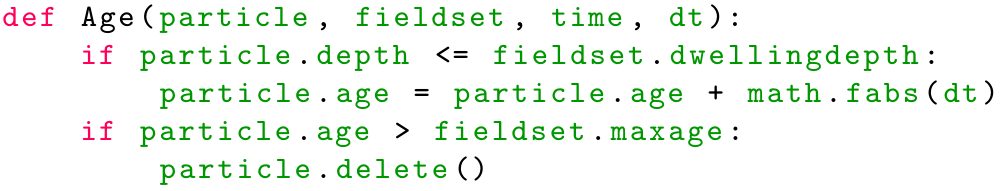}

The third kernel, which samples the temperature, can be written as

\includegraphics[scale=1.0]{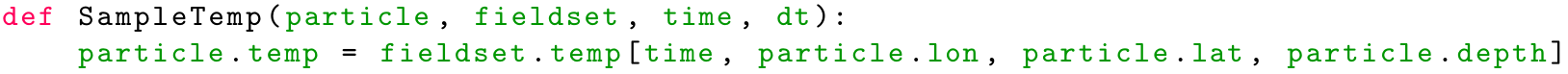}

These three kernels are then concatenated with the \verb|AdvectionRK4_3D| kernel as

\includegraphics[scale=1.0]{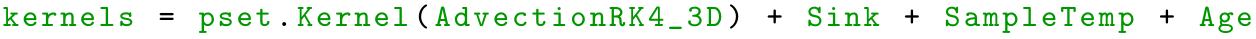}

Where at least one of the kernels needs to be cast into a \verb|Kernel| object
for the overloading of the + operator as a kernel concatenator to work.

Finally, the kernel that deletes a particle if it reaches one of the lateral
boundaries and which will be invoked through the error recovery execution is

\includegraphics[scale=1.0]{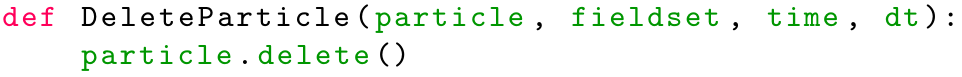}

\subsection{Executing the particle set}\label{sec:example_executing}

The \verb|ParticleSet| can now be integrated with a call to
\verb|pset.execute()|. This method requires as input the list of
kernels, the starttime of the execution loop, the runtime of the
execution loop, the Runge Kutta integration timestep (here taken to be 5
minutes),  the interval at which output is written (here once per day),
and the recovery kernel that gets called when a Particle crosses the
boundary of the regional domain.

As mentioned in section \ref{sec:example_fieldset}, only three
timeslices are held in memory at any one time. The loading of new fields
is controlled by the \verb|fieldset.advancetime()| method, which
replaces the oldest timeslice with a new one (held in this case in
\verb|[snapshots[s]]|). This also means that the executing of the
ParticleSet has to be done within a loop: 

\includegraphics[scale=1.0]{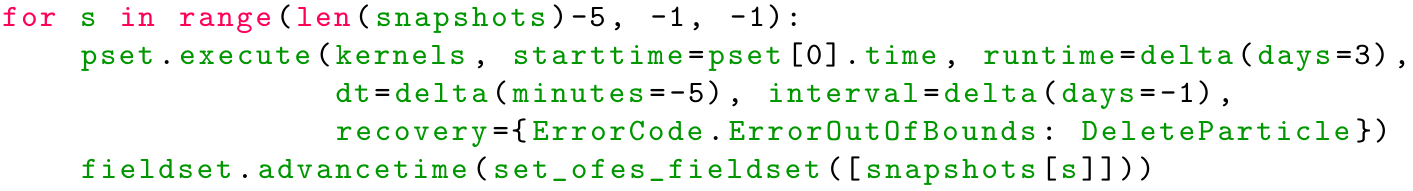}

There is another reason to call the \verb|pset.execute| method within a loop: it allows
for a new particle to be released every three days (the frequency with which
hydrodynamic data is available). This happens within the for-loop through a call
to 

\includegraphics[scale=1.0]{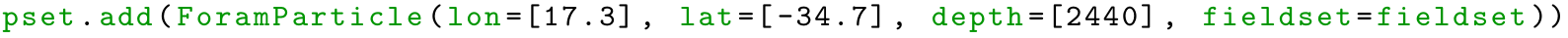}

\subsection{Saving and plotting the output}

The Parcels framework allows for storing of the locations of the
particle to disk on-the-fly in NetCDF files, following the Discrete
Sampling Geometries section of
\url{http://cfconventions.org/cf-conventions/v1.6.0/cf-conventions.html\
#discrete-sampling-geometries}, and is hence CF-1.6-compliant. Storing
of the particle trajectories and properties such as age and along-track
temperature happens in the for-loop through calls to

\includegraphics[scale=1.0]{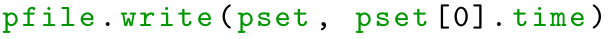}

Since particles are continually added to and deleted from the ParticleSet, the
\verb|ParticleFile| needs to be stored in `indexed' format, where for each
variable all particle states are written in one long vector.

\includegraphics[scale=1.0]{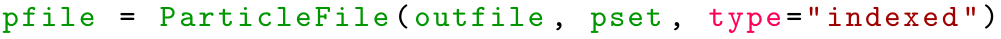}

These long vectors in Indexed format, however, are not very easy to work
with, so Parcels provides the utility script \verb|convert_IndexedOutputToArray|
to convert an Indexed NetCDF file to array format.

\begin{figure}[t] \centering
\includegraphics[width=\linewidth]{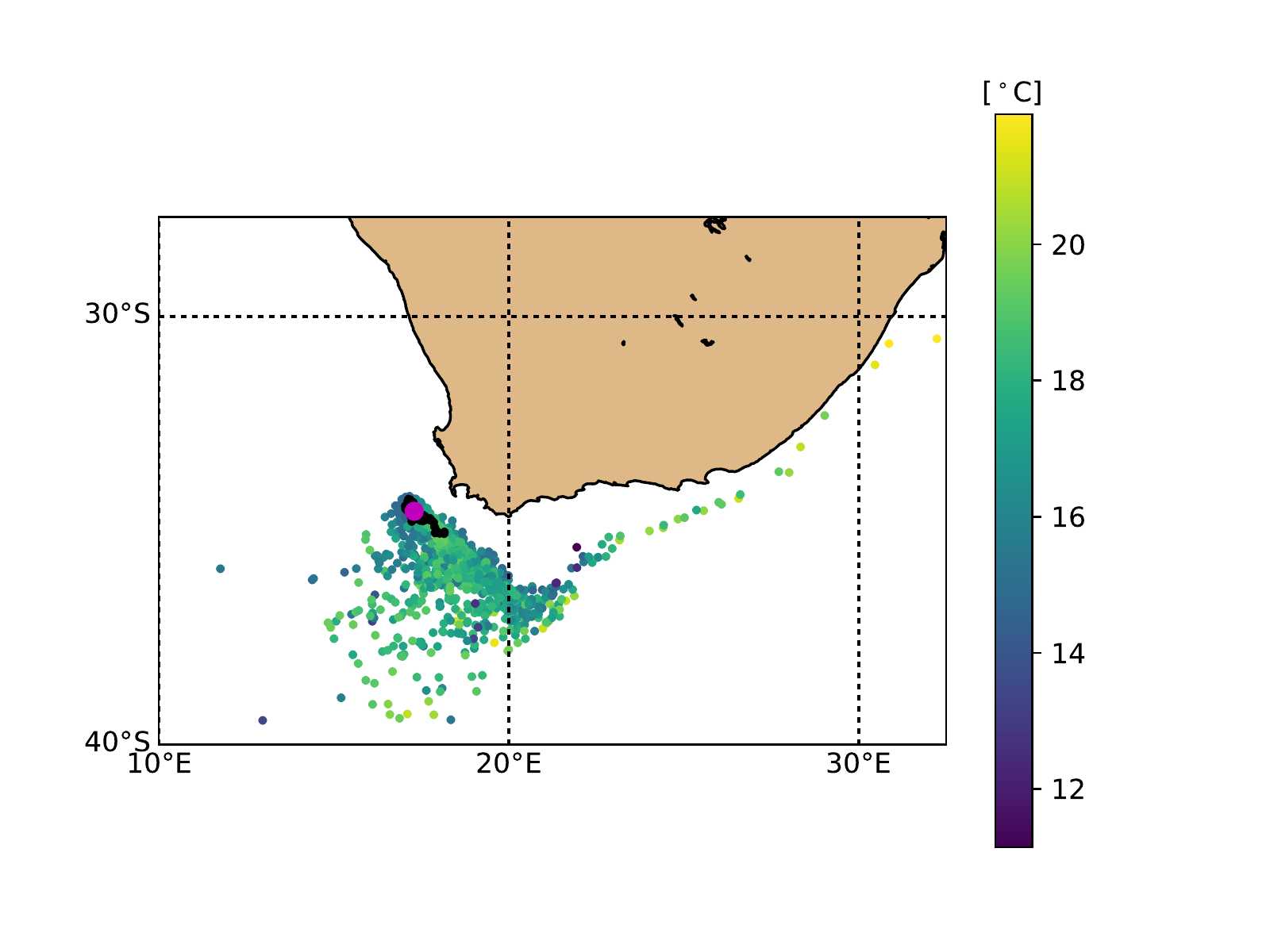}
\caption{Footprints of virtual foraminifera ending up on the ocean floor just
off Cape Town in the Agulhas region. This experiment is a Parcels implementation
of the study described in \citet{vanSebille:2015ed}, and this figure can be
compared to Fig.~1b in that paper. The magenta dot is the location of the
sediment core, from which virtual particles are first tracked back until they
reach their 50m dwelling depth (black dots), and then further tracked back for
their 30-day lifespan. Temperatures (in degrees Celcius) are recorded each day
throughout their lifespan and shown as colours. The code for this experiment and
plotting is available at \url{https://doi.org/10.5281/zenodo.823994.}}
\label{example_corefootprint}
\end{figure}

The particle trajectories can then be plotted using the \verb|matplotlib| and
\verb|Basemap| libraries, see Fig.~\ref{example_corefootprint}. This figure
shows the temperature recorded on each day during the lifespan of all virtual
particles. It highlights that foraminifera that end up on the ocean
floor off Cape Town travel hundreds to thousands of kilometers during their
lifespan, and that while some originate from the Agulhas Current as far north as
27$^\circ$S, others originate from the much colder Southern Ocean south of
40$^\circ$S.

\section{Model evaluation}\label{sec:eval}
Evaluation of a code-base's accuracy and performance is a key component of its
validation and roll-out. For this Parcels v0.9, performance and speed are not
a priority; these will be the focus for the v1.0 release (see also
Sect.~\ref{sec:outlook}). Instead, while developing Parcels v0.9 we have
concentrated on accuracy.

\subsection{Unit tests and continuous integration}

Following best practices in software engineering, we have incorporated Unit
Testing and Continuous Integration into the development cycle of Parcels. Every
push of code changes to github automatically triggers a validation of the
entire code base (an important component of the Continuous Integration
paradigm), through the \url{travis-ci.org} web service.

The validation of the code base is done through so-called unit tests; small
snippets of code that test individual components of the codebase. Parcels v0.9
has over 150 of these unit tests, which check the integrity and consistency of
the codebase. Where relevant, these unit tests are run in both Scipy and JIT
mode, to test both modes of executing the kernels.

The following Python snippet shows a typical example of a unit test for Parcels
(as included in the \verb|test_particle_sets.py| file). It performs the test
that \verb|Particles| in a \verb|ParticleSet| indeed get their assigned
longitudes and latitudes. While this may seem a trivial test, these kinds of
unit tests can help prevent bugs.

\includegraphics[scale=1.0]{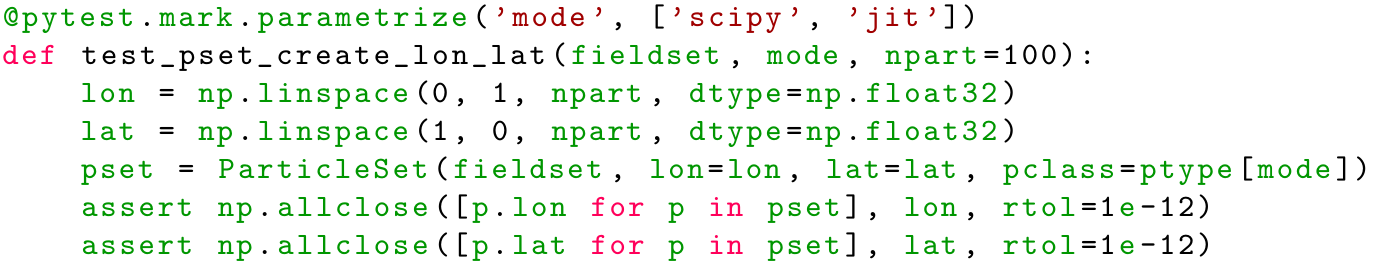}

Ideally, the full set of unit tests means that no change of the code can ever
break another part of the code, since some of the unit tests would then fail. Of
course, in reality the completeness of the unit tests can never be guaranteed,
but during Parcels development we have attempted to provide unit tests for a
broad spectrum of the Parcels functionality and code.

\subsection{Idealised and analytic test cases}\label{sec:eval-analytic}

Following the list of standard tests of particle tools, as described in Sec.~6
of \citet{vansebille2017}, we have validated the accuracy of Parcels v0.9
against seven idealised and analytical test cases. In this section we will
describe the results in detail. All test cases are run with
Runge-Kutta4 integration and in JIT mode. In each case, the hydrodynamic
velocities are generated within the Python scripts and converted
directly to a \verb|FieldSet| (i.e. without first storing these fields
in NetCDF format).The Python code for all testcases is
available at \url{https://doi.org/10.5281/zenodo.823994}.

\subsubsection{Radial rotation with known period}

The first test case is that of a simple counter-clockwise solid-body
rotation with a period of 24 hours. Velocities are defined on a $(20
\times 20)$ km Arakawa A-grid centered at the origin with a $100$ m
horizontal resolution. Solid-body radial velocities $(u, v) = \left(-\omega r
\sin(\phi), \omega r \cos(\phi)\right)$, with $r$ and $\phi$ the radius and
angle from the origin and $\omega=2\pi/86,400$ s the angular frequency,
are then computed on that grid.

Four particles are started at $x=0$ km and $y = (1000, 2000, 3000,
4000)$ km and then advected for 24 hours, using an RK4 timestep of 5
minutes, and with particle positions stored every hour
(Fig.~\ref{fig:evaluations}a). All four particles indeed follow the flow
for the full circle. The maximum distance error after this 24 hours
advection is less than 3mm, on path lengths of more than 5km.

\subsubsection{Longitudinal shear flow }

\begin{figure}[t] \centering
  \includegraphics[width=\linewidth]{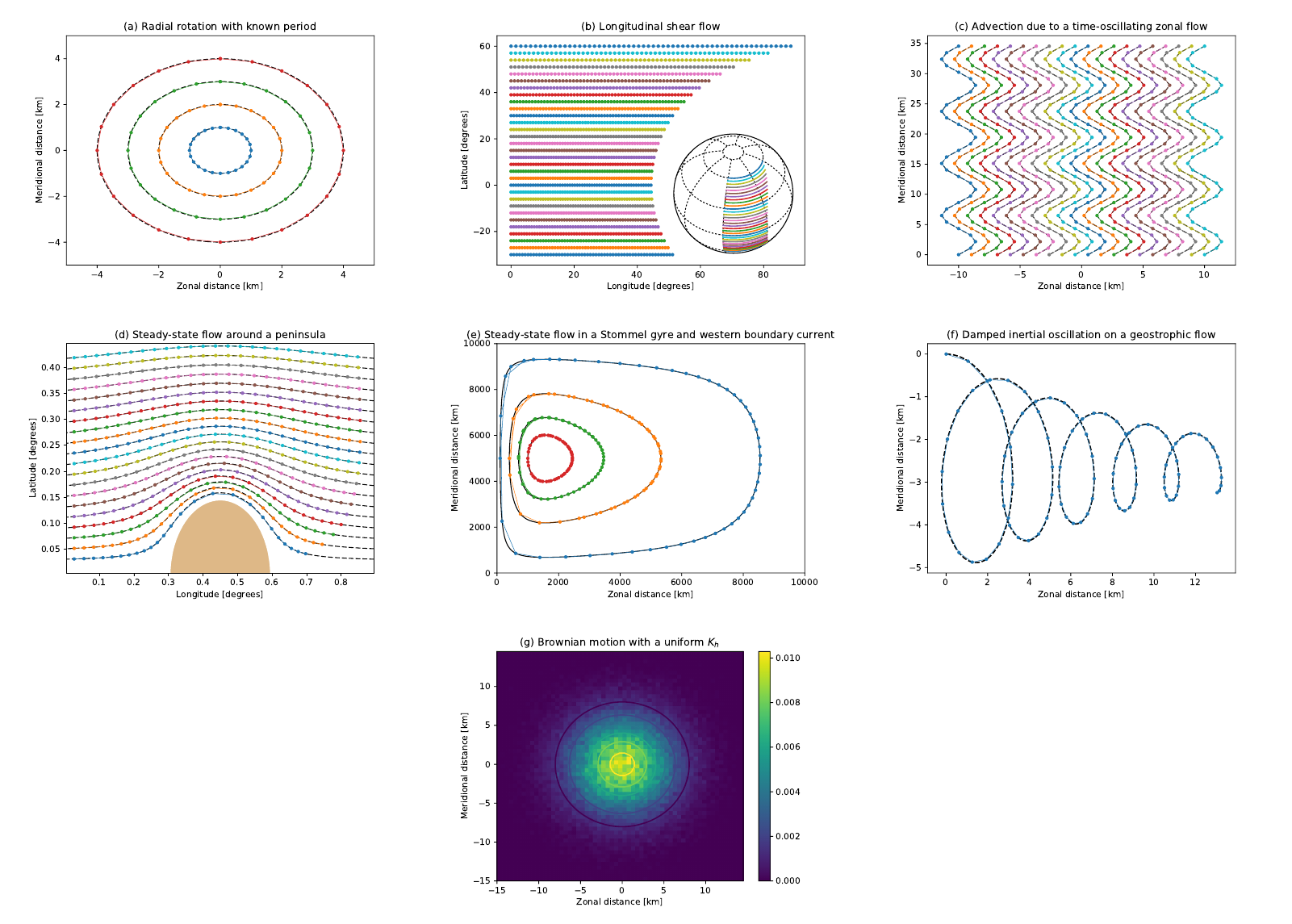}
  \caption{Evaluation of trajectory accuracy in Parcels v0.9, following
  the seven idealised and analytical test case described in in Sect.~6
  of \citet{vansebille2017}: (a) radial rotation with known period; (b)
  longitudinal shear flow; (c) advection due to a time-oscillating zonal
  flow; (d) steady-state flow around a peninsula; (e) steady-state flow
  in a Stommel gyre and western boundary current; (f) damped inertial
  oscillation on a geostrophic flow; and (g) Brownian motion with a
  uniform $K_h$. In the upper six panels, the coloured lines are the
  particle trajectories and the black dashed lines are the analytical
  solutions. In panel (g), the colouring shows the density of particles,
  and the contours show the probability density function of the
  equivalent analytical solution (a two-dimensional Gaussian).}
  \label{fig:evaluations}
\end{figure}

The second test case tests the ability of the Parcels code to convert
between spherical longitude/latitude space and local flat Euclidian space. When defining
a \verb|FieldSet| on a spherical mesh, Parcels automatically performs this
conversion under the hood. To test its accuracy, an idealised flow on a
sphere at $1^\circ$ horizontal resolution is created, 
with a uniform zonal velocity of 1 m/s and no
meridional velocity. A total of 31 particles are then released on a
north-south line, with a meridional spacing of 3$^\circ$. These
particles are advected for 57 days, using an RK4 timestep of 5 minutes
and output saved every day (Fig.~\ref{fig:evaluations}b). The main
panel shows trajectories in planar projection, with the inset showing
the same trajectories in orthographic projection.

At a speed of 1 m/s, the particles travel 4.9$\cdot 10^6$ m in the 57 days. At
the equator, this amounts to almost 45$^\circ$ of longitude, but
because of the cosine-dependence of zonal distance with latitude, particles
closer to the poles travel farther in degrees (main panel in
Fig.~\ref{fig:evaluations}b). The inset of Fig.~\ref{fig:evaluations}b,
nevertheless, shows that in an orthographic projection, all particles travel 
the same distance.

\subsubsection{Advection due to a time-oscillating zonal flow}

The third test case tests the ability of Parcels to cope with simple
time-varying flow. The flow in this case is a uniform meridional flow of $v = A
=0.1$ m/s, and an oscillating zonal flow with $u(t)= A \cos(\omega t)$ where
$\omega = 2\pi / T$ and the period is $T=1$ day. 
The time resolution of the \verb|FieldSet| is 5 minutes, and since the
flow is constant in space there are only two grid cells in each of the
horizontal directions.
A total of 20 particles are
then released on a zonal line at $y=0$ km and advected for 4 days, using an RK4
timestep of 5 minutes and storing output every 3 hours
(Fig.~\ref{fig:evaluations}c).

The analytical flow for the paths of these particles is $y(t) = At$ and $x(t) =
x_0 + A/\omega\sin(\omega t)$ where $\omega = 2\pi / T$ and $x_0$ is the  zonal
start location of the particle. Indeed, all particles follow these analytical
pathways very closely (Fig.~\ref{fig:evaluations}c), with largest positional
errors after 4 days being 6 cm in the zonal direction and 4 mm in the meridional
direction.

\subsubsection{Steady-state flow around a peninsula}\label{sec:eval-peninsula}

The test case of steady-state flow around a peninsula follows a description by
\citet{Adlandsvik:2009tx} and was also used as a validation test case in the article
describing the Connectivity Modeling System \citep{Paris:2013fs}. Starting from
the analytical expression for a streamfunction $\Psi$ of a steady-state flow
around a peninsula, analytical expressions of the zonal and meridional component
of velocity are solved on a $(1^\circ \times 0.5^\circ)$ Arakawa A-grid 
at $1/100^\circ$ horizontal resolution. A set of 20 particles is
seeded just off the western edge of the domain, and then advected with the flow
for 24 hours using an RK4 timestep of 5 minutes and particle positions
stored every hour (Fig.~\ref{fig:evaluations}d, where the brown semi-circle is
the peninsula).

Since the particles should follow streamlines, a comparison of the interpolated
streamfunction value at $t=24$ hours to that at $t=0$ hours gives an estimate of
the error. The largest error is 0.008 m$^2$/s, which corresponds to a positional
error of $10^{-5}$ degrees, or 1 meter. Indeed, Fig.~\ref{fig:evaluations}d shows
that the particle trajectories closely follow the dashed streamlines.

\subsubsection{Steady-state flow in a Stommel gyre and western boundary current}

The test case of the Stommel gyre follows a description in
\citet{Fabbroni:2009uw}, and provides an analytical solution to the
streamfunction field of a Stommel gyre and western boundary current. Here, we
compute the meridional and zonal central derivatives of this streamfunction
field to generate zonal and meridional velocities, respectively, on a $(10,000 \times
 10,000)$ km Arakawa A-grid at 50 km horizontal resolution. 
A set of four particles is seeded on a line crossing the western
boundary, at $y=$ 5,000 km, and then advected for 50 days with an RK4 timestep of
5 minutes and the particle positions stored every 24 hours
(Fig.~\ref{fig:evaluations}e).

Since the particles should follow streamlines, the deviation of particles from
the streamlines is a measure of the accuracy of the method.
Fig.~\ref{fig:evaluations}e shows that all three particles stay close to
their streamline throughout the 50 day advection period. The largest error is
0.05 m$^2$/s, which corresponds to a positional error of less than 5 km.

\subsubsection{Damped inertial oscillation on a geostrophic flow}

The test case of a damped inertial oscillation on a geostrophic flow follows
\citet{Fabbroni:2009uw} and \citet{doos:2013}. In this test case, the velocity varies
over the entire domain, following an analytical time-dependent equation. Here,
we use a time resolution of 5 minutes for the velocity field. A particle is then
seeded at the origin and advected for four days, with a RK4 timestep of 5
minutes and output stored every hour (Fig.~\ref{fig:evaluations}f). After four
days of advection, the positional error of the particle, as compared to the
analytical solution, is less than 5 cm.

\subsubsection{Brownian motion with uniform $K_h$}\label{sec:eval-brownian}

The test case of Brownian motion with uniform $K_h$ tests for the accuracy and
implementation of the random number generator. Here, a total of 100,000
particles are seeded at the origin of a $(60 \times 60)$ km grid
centered around the origin with zero velocities, and then diffused using
a normal variate
random number distribution with $K_h = 100$ m$^2$/s. The particles are diffused
for 1 day with a timestep of 5 minutes (Fig.~\ref{fig:evaluations}g). The
two-dimensional normalised histogram agrees very well with the analytical
solution of this Brownian motion: a two-dimensional Gaussian with a mean at the
origin and standard deviation of $\sigma = \sqrt{2K_h\Delta t}= 4.16$ km.

\section{Future outlook}\label{sec:outlook}
As mentioned before, Parcels v0.9 is a prototype. The core
contributions of this paper are both the API, as well as the design philosophy which
enables a wide range of valuable future improvements of the framework.
Below, we discuss some of the conceptual ideas for these planned
improvements.

\subsection{Performance optimisation}

The primary performance optimisation in version 0.9 of
Parcels is the automated generation of C kernel code to
allow inlining of field evaluation routines. However, several future
optimisations have been planned during the design of the code,
based around considerations for irregular data processing. Since
dynamic addition and deletion of particles is a common feature of many
oceanographic use cases, no assumptions about data layout or iteration
protocol have been made in the high-level API of particle sets,
allowing more optimised implementations in the future. The use of
dynamic code generation at runtime also enables further automated
specialisation of kernel code, while allowing us to define a clear
initial interface for kernel customisation.

In addition to optimising the execution of particle kernels, the
extensive interaction with hydrodynamic field data constitutes a
considerable cost of the overall computation - a cost that is likely
to dominate overall execution if large sets of hydrodynamic field data
are to be read from files. Multiple potential approaches can be
considered in future versions of Lagrangian particle tracking codes:

\begin{itemize}
\item Directly coupled (online) runs within the host OGCM
  can completely avoid the bandwidth bottlenecks imposed by reading
  dense field data from disk, at the expense of additional computation.
  For simulations at local scales with a high particle density, this
  trade-off might prove beneficial, for example for regional studies
  on marine ecology.
\item For global-scale models that require offline hydrodynamic field
  data but feature a low particle density with high localization, the
  total volume of data read from disk might be drastically reduced by
  explicitly prefetching local subsets of field data based on particle
  locations. Such a mechanism would require the use of additional
  geospatial indexing methods, for example via octrees or
  r-trees~\citep{Isaac2015,Schubert2013}, that decompose the grid into
  individual sub-regions and provide fast indexing methods. Using
  explicit prefetch directives in the dynamic execution loop might
  also enable overlapping of asynchronous file reads with effective
  computation to further amortize file I/O overheads.
\end{itemize}

The modularity of Parcels' internal abstractions, as well as the
composability of kernels and the flexibility provided by the dynamic
execution loop should facilitate extensive experimentation and
exploration with such advanced optimization techniques, without
the need for users to change any high-level algorithmic definitions.
The use of advanced data handling and task-scheduling libraries, such
as Dask~\citep{Rocklin2015} or Xarray~\citep{Hoyer2017}, might also be
utilised to quickly achieve efficient out-of-core data management in
Parcels.

\subsubsection{Towards parallellisation}

The current version of Parcels is not in itself parallel due to two
restrictions:

\begin{itemize}
\item The primary input format of field data in the v0.9 prototype is
  NetCDF-based field data, so that parallellisation requires an
  explicit domain decomposition and a parallel file reader. The
  current version of the \texttt{netcdf} Python package does not provide
  these features. Alternative implementations of the NetCDF file format,
  such as Xarray, might be
  leveraged in future versions of Parcels to provide parallel data
  management.

\item Exchanging particle information between parallel processors is
  currently not supported, although it is deemed a critical feature
  for the next release (v1.0).
\end{itemize}

%While the primary focus of the current v0.9 release is on API and
%functionality, the most significant addition for future releases
%will be the addition of multi-processor parallelism.

\subsection{Community building and kernel sharing}

One of the key ideas between the development of Parcels is for it to be a
flexible and extendable codebase, where particle behaviour can easily be
customised. The worked out example in Sect.~\ref{sec:example} shows that many
types of behaviour (sinking, aging, etc) can be coded in a few lines of Python
code.

The customisability of Parcels enables a multitude of oceanographic
modelling, from water parcels to plankton to plastic litter to fish. We
therefore envision an active community of Parcels users who share and
discuss kernel development. We encourage anyone who wishes to share
their custom kernels to upload them onto github, and we will provide a
properly referenced library of user-contributed kernels for others to
reuse on \url{oceanparcels.org}.

\subsection{Towards runtime integration with OGCMs}

Although the current version of Parcels primarily uses off-line field
data, the overall design of the particle exectuion engine is designed
to be compatible with a variety of OGCMs for directly coupled (at-runtime)
simulations. In particular, the current \texttt{Field} interface can
easily be extended to provide interpolation routines for various types
of field data, for example based on unstrctured meshes, while the
primary particle update loop provides a mechanism for host models to
dictate a model timestep size that varies from that of the particle
update. Moreover, the explicit generation of C code allows Parcels
kernel code to be easily injected into existing ocean modelling
frameworks, while the provision of error-recovery kernels can guarantee
progression of the coupled model.

\conclusions\label{sec:conclusions}

Here, we have introduced a new framework for Lagrangian ocean analysis
that focusses on customisability, flexibility and ease-of-use. This v0.9
of Parcels is very much a prototype, providing a proof-of-concept of the
API and showcasing how it can be used to create high-level Python code
for full-fledged scientific experiments. We also assess the accuracy of
the current implementation, with the idea to provide a benchmark for
future versions. Future development will focus on increasing efficiency
of the framework, and also towards providing easy tools to port the
generated C-code of Parcels experiments to at-runtime integration within
OGCMs.

\section{Code availability}
The code for Parcels is licensed under the MIT license and is available through
github at \url{github.com/OceanParcels/parcels}. The version 0.9 described here
is archived at Zenodo at \url{https://doi.org/10.5281/zenodo.823562}. More
information is available on the project webpage at \url{oceanparcels.org}.

\authorcontribution{ML and EvS developed the code and wrote the manuscript jointly.}

\competinginterests{The authors declare no competing interests}

\begin{acknowledgements}
The initial ideas for the Parcels framework were the result of very
fruitful discussions with the attendees of the ``Future of Lagrangian
Ocean Modelling'' workshop, held at Imperial College London, UK, in
September 2015. Funding for this workshop was provided through an EPSRC
Institutional Sponsorship grant to EvS under reference number
EP/N50869X/1. EvS is supported through funding from the European
Research Council (ERC) under the European Union's Horizon 2020 research
and innovation programme (grant agreement No 715386). The OFES
simulation was conducted on the Earth Simulator under the support of
JAMSTEC. We thank Joe Scutt-Phillips, Ronan McAdam, Joel Kronberg,
Thomas Stokes, Nathaniel Tarshish, Michael Hart-Davis, Birgit Sutzl, Ben
Snowball, Samuel Wetherell and David Ham for their support in testing
and developing aspects of the Parcels code.
\end{acknowledgements}

\bibliographystyle{copernicus}
\bibliography{biblio.bib}

\clearpage
\clearpage

\clearpage

\clearpage

\end{document}